\documentclass[aps,preprintnumbers,nofootinbib,floatfix,fleqn]{revtex4}
\usepackage{enumitem}  
\usepackage{amssymb,bm,graphicx}
\usepackage{amsmath,bbold,ulem}
\usepackage{dsfont}
\usepackage{slashed}
\usepackage[dvipsnames]{xcolor}
\usepackage[utf8]{inputenc}
\usepackage{wasysym}

\newcommand{\be}{\begin{equation}}
\newcommand{\ee}{\end{equation}}
\newcommand{\ba}{\begin{eqnarray}}
\newcommand{\ea}{\end{eqnarray}}
\newcommand{\bsub}{\begin{subequations}}
\newcommand{\esub}{\end{subequations}}

\allowdisplaybreaks

\begin{document}

\title{Quark-model relations among TMDs in the parton model}

\author{
F.~Aslan$^1$,
S.~Bastami$^2$,
A.~Mahabir$^1$,
A.~Tandogan$^{1,3}$,
P.~Schweitzer$^1$}
  \affiliation{
  $^1$
  Department of Physics, University of Connecticut, 
  Storrs, CT 06269, U.S.A.\\
  $^2$  
  Department of Physics, Kennesaw State University,
  30144 Kennesaw, GA, U.S.A.\\
  $^3$  
  Department of Physics, University of Connecticut, 
  Hartford, CT 06103, U.S.A.}

\begin{abstract}
The covariant parton model (CPM) is a consequent
application of the parton model concept to the 
nucleon structure.
In this model, there is a choice to put quarks 
either in a pure-spin state or in a mixed-spin state.
We show that the mixed-spin version of the CPM does
not support the quark-model relations among 
transverse momentum dependent parton distributions
(TMDs) which were shown to hold in a large
class of quark models. One can enforce the 
quark-model relations to be valid in the CPM  
by imposing a condition which is equivalent
to putting the quarks in a pure-spin state.
This gives a complementary perspective on the
connection of the pure- and mixed-spin state 
CPM versions, and provides a fresh view on the
question whether the quark-model relations 
could be realized in QCD as ``approximate relations'' with some useful numerical accuracy.
\end{abstract}

\maketitle

\section{Introduction}
\label{Sec-1:intro}

TMDs entail the description of the nucleon
structure in deep-inelastic scattering~(DIS)
processes when in the final state one detects
an adequate transverse momentum which is small 
compared to the hard scale $Q$ of the
process~\cite{Collins-book}. 
For the understanding of the nonperturbative
properties of TMDs, quark models play an
important role in two ways. 
First, undistracted by technical
complexities inherent in a full QCD
treatment, in models one may investigate
in a simpler theoretical framework the
significance of a specific physical aspect
and gain in this way valuable insights.
Second, in situations where some of the TMDs 
are still not yet well known, results from models 
may be helpful to interpret first data or give
useful estimates for counting rates in future 
experiments such as the Electron-Ion Collider
\cite{Accardi:2012qut}. 
In this way, models complement phenomenology 
and lattice QCD studies.

In this work, we will study the covariant 
parton model (CPM) which is based on Feynman's
parton model concept 
\cite{Feynman:1969ej,Feynman:1973xc}. 
The latter played a historically important role
for the interpretation of DIS processes and
establishing QCD and can, in a certain sense, 
be viewed as a ``zeroth order approximation'' 
to QCD \cite{Ellis:1978ty,Collins-book}.
The parton model provides often an effective
first step towards an understanding of QCD
processes.
For instance, the ``generalized parton model'' 
of Refs.~\cite{Anselmino:1994tv,Anselmino:1999pw,
Anselmino:2005nn,Anselmino:2005sh,Anselmino:2011ch}
helped to pave the way to modern TMD
phenomenology. The exploration of the 
parton model concept for the sake of 
studying TMDs and their nonperturbative
properties was carried out in 
Refs.~\cite{
Zavada:1996kp,Zavada:2001bq,Zavada:2002uz,
Efremov:2004tz,Zavada:2007ww,Efremov:2009ze,
Zavada:2009ska,Efremov:2010mt,Zavada:2011cv,
Zavada:2013ola,Zavada:2015gaa,Zavada:2019yom,
Bastami:2020rxn,DAlesio:2009cps,Aslan:2022wqc}. 
Further applications of the parton model 
concept can be found in
Refs.~\cite{Blumlein:1996tp,Blumlein:1996vs,
Blumlein:1998nv,Roberts:1996ub,Jackson:1989ph}.

Due to the absence of interactions, the description 
of the nucleon structure in the parton model is 
particularly lucid and the TMDs are described in 
terms of covariant functions depending on the 
variable $P\cdot k$ where $P^\mu$ denotes the 
nucleon momentum and $k^\mu$ quark momentum.
Despite the simplicity of the model, 
there was an interesting puzzle.
One group claimed that the description of TMDs
requires two independent covariant functions
\cite{
Efremov:2009ze,Zavada:2009ska,Efremov:2010mt,
Zavada:2011cv,Zavada:2013ola,Zavada:2015gaa,
Zavada:2019yom,Bastami:2020rxn},
while the other group claimed that one needs
three independent covariant functions  
\cite{DAlesio:2009cps}. 
This puzzle was resolved recently by showing 
that the results of the two groups are 
equivalent except for the treatment of the 
quark polarization state \cite{Aslan:2022wqc}.
In \cite{Zavada:2001bq,Zavada:2002uz,
Efremov:2004tz,Zavada:2007ww,Efremov:2009ze,
Zavada:2009ska,Efremov:2010mt,Zavada:2011cv,
Zavada:2013ola,Zavada:2015gaa,Zavada:2019yom,
Bastami:2020rxn}
the quarks were chosen to be in a pure-spin
state, while in \cite{DAlesio:2009cps} they 
were (implicitly) assumed to be in a 
mixed-spin state. Other than that, the
results of the two groups are equivalent
\cite{Aslan:2022wqc}.

Here we will take a different point 
of view as compared to Ref.~\cite{Aslan:2022wqc}
where the focus was on technical aspects of the 
quark correlator.
In this work the starting point is the quark model
aspect of the approach: the CPM is after all a quark
model, i.e.\ a model 
without gauge field degrees of freedom. In several
models of such type it was observed that certain
relations exist between different TMDs to which 
we shall refer as quark model relations (QMRs). 
Not all quark models support the QMRs, 
but it is worth stressing that a wide
class of very different models does.

The goal of this work is to investigate whether
the TMDs in the mixed-spin state version of the 
CPM \cite{DAlesio:2009cps,Aslan:2022wqc} obey
the QMRs. We shall see that this is not the 
case and show that imposing the validity of
the QMRs in this model yields the same 
condition as when one chooses the quarks
to be in pure-spin state. In other words,
if one starts with the mixed-spin state version 
of the CPM and demands the model to comply with
the QMRs supported in other quark models, then one
must introduce the pure-spin state model.

Our study is insightful in two ways. First, 
it gives insights on the CPM and its relation 
to other quark models.
Second, it opens a new perspective on QMRs
and may shed light on the question whether they 
could hold in QCD as approximate relations with 
a potentially useful numerical accuracy 
in some range of $x$ and $k_T$.

The structure of this work is as follows. 
In Sec.~\ref{Sec-2:QMRs} 
    we present the QMRs and briefly discuss
    their understanding 
    within quark models.
In Sec.~\ref{Sec-3:CPM-mixed-spin} 
    we review the CPM and present the results for TMDs in the mixed-spin state
    version of the CPM.
In Sec.~\ref{Sec-4:imposing-QMRs}
    we investigate the linear and
    non-linear QMRs in the CPM.
In Sec.~\ref{Sec-5:discussion}
    we discuss the physical implications of our 
    findings, and
in Sec.~\ref{Sec-6:conclusions} 
    we draw conclusions and give an
    outlook for future studies.
    

\section{Quark model relations among TMD\lowercase{s}}
\label{Sec-2:QMRs}

In contrast to QCD, in quark models relations among different TMDs
can exist due to the simpler model dynamics or due to model symmetries.
Some of these relations such as, e.g, the quark-model Lorentz-invariance 
relations (qLIRs), are generic in the sense that they hold in quark 
models which respect Lorentz symmetry and contain no gauge field 
degrees of freedom \cite{Tangerman:1994eh,Mulders:1995dh,Boer:1997nt},  
but are not valid in QCD \cite{Kundu:2001pk,Goeke:2003az,Goeke:2005hb}. 
We quote here only one qLIR, namely
\be
      h_T^{ q}(x,k_T)-h_T^{\perp  q}(x,k_T)
      = h_{1L}^{\perp  q}(x,k_T) \label{Eq:qLIR-example}\,,
\ee
on which it will be instructive to follow up below.
A discussion of other qLIRs can be found for instance in
Ref.~\cite{Metz:2008ib}. We remark that the notation in
(\ref{Eq:qLIR-example}) and throughout this work is
$k_T = |\vec{k}_T|$ and $k_T^2$ will always denote $|\vec{k}_T|^2$.

The main focus of this work is another set of relations which have been
observed in several very different quark models. These relations, to which
we will refer to in the following as quark model relations (QMRs), are 
given by
\bsub\label{Eq:quark-model-relations-linear}
\ba
      g_{1T}^{\perp  q}(x,k_T) 
      &=& - h_{1L}^{\perp  q}(x,k_T), 
      \label{Eq:qm-rel-1}\\
      g_{ T}^{\perp  q}(x,k_T) 
      &=& - h_{1T}^{\perp  q}(x,k_T), 
      \label{Eq:qm-rel-2}\\
      g_{ L}^{\perp  q}(x,k_T) 
      &=& - h_{ T}^{ q}(x,k_T), 
      \label{Eq:qm-rel-3}\\
      g_1^{ q}(x,k_T)-h_1^{ q}(x,k_T) 
      &=& h_{1T}^{\perp(1) q}(x,k_T), 
      \label{Eq:qm-rel-4}\\
      g_T^{ q}(x,k_T)-h_L^{ q}(x,k_T) 
      &=& h_{1T}^{\perp(1) q}(x,k_T), 
      \label{Eq:qm-rel-5}
\ea\esub
In addition to the linear QMRs (\ref{Eq:quark-model-relations-linear})
also two nonlinear QMRs have been found which are given by
\bsub\label{Eq:quark-model-relations-nonlinear}
\ba
    \frac12\biggl[h_{1L}^{\perp  q}(x,k_T)\biggr]^2
    &=&     
    -\,h_1^{ q}(x,k_T)\,h_{1T}^{\perp  q}(x,k_T)
    \,,
    \label{Eq:qm-rel-7}\\
    \frac12\biggl[g_{1T}^{\perp  q}(x,k_T)\biggr]^2
    &=&
    g_{1T}^{\perp q}(x,k_T)\,g_{L}^{\perp q}(x,k_T)+
    g_T^{ q}(x,k_T)\,g_{T}^{\perp q}(x,k_T)
    \,.
    \label{Eq:qm-rel-8}
\ea
\esub
The transverse moment of a TMD is defined as 
\be
     h_{1T}^{\perp(1)q}(x,k_T) = \frac{k_T^2}{2M^2}\,h_{1T}^{\perp q}(x,k_T)\,.
\ee

The relations (\ref{Eq:quark-model-relations-linear},
\ref{Eq:quark-model-relations-nonlinear}) hold in a 
wide class of quark models which are based on very
different model concepts including the spectator 
model, bag model, or light-front constituent quark model 
\cite{Jakob:1997wg,Avakian:2007mv,Avakian:2008dz,Avakian:2009jt,She:2009jq,Avakian:2010br,Pasquini:2008ax,
Lorce:2011zta}.
The QMRs (\ref{Eq:qm-rel-1},~\ref{Eq:qm-rel-4},~\ref{Eq:qm-rel-7})
involving twist-2 TMDs were shown to arise from a certain rotational 
symmetry of the model lightcone wave functions
\cite{Lorce:2011zta}. This symmetry is effectively
present in many models including, e.g., the 
pure-spin version of the CPM
\cite{Efremov:2009ze,Bastami:2020rxn}.

It is important to remark that not all models 
support the QMRs. For instance, certain 
spectator model variants, where (to allow more
flexible modeling) a larger number of free model
parameters was introduced \cite{Bacchetta:2008af}, 
do not support QMRs. Another example is
quark-target model \cite{Meissner:2007rx} where
the presence of gluon degrees of freedom spoils QMRs.

\section{Quark correlator and TMD\lowercase{s} 
in CPM}
\label{Sec-3:CPM-mixed-spin}

In this section, we first review the general structure 
of the quark correlator in quark models and then discuss
the specific results for the correlator and TMDs in the
CPM briefly commenting on the two versions of this model.

\subsection{Quark correlator in a generic quark model}

In a theory without explicit gauge degrees of freedom the quark 
correlator for the nucleon is defined as follows
\ba\label{Eq:correlator-q}
	\Phi_{ij}^q(k,P,S) &=& 
	\int \frac{\mathrm{d}^4z}{(2\pi)^4}\;\mathrm{e}^{i k z}\,
    \langle N|\,\overline{\Psi}_j^{\,q}(0)\;\Psi_i^q(z)\,
    |N\rangle\,,
\ea
where $k^\mu$ is the quark 4-momentum, and $P^\mu$ and $S^\mu$ are 
the nucleon 4-momentum and polarization vectors satisfying $P^2=M^2$, 
$S^2=-1$, $P\cdot S=0$. In quark models, the Lorentz-structure of the 
correlator (\ref{Eq:correlator-q}) is described in terms of $k^\mu$, 
$P^\mu$, $S^\mu$ as follows (we use the convention $\varepsilon^{0123} = 1$
and assume a covariant normalization of nucleon states) \cite{Boer:1997nt} 
\begin{align}
    \label{Eq:correlator-decompose}
     \Phi^q(k,P,S) 
     &=
     MA_1^q + \slashed{P} A_2^q + \slashed{k}A_3^q
     + \frac{i}{2M} \; [\slashed{P},\slashed{k}] \,A_4^q
     + i (k\cdot S) \gamma_5 A_5^q 
     + M\slashed{S} \gamma_5 A_6^q
     + \frac{(k\cdot S)}{M} \slashed{P} \gamma_5 A_7^q
     + \frac{(k\cdot S)}{M} \slashed{k} \gamma_5 A_8^q
 \nonumber\\
   & + \frac{[\slashed{P},\slashed{S}]}{2}\gamma_5 A_9^q 
     + \frac{[\slashed{k},\slashed{S}]}{2} \gamma_5  A_{10}^q 
     + \frac{(k\cdot S)}{2M^2} [\slashed{P},\slashed{k}] \gamma_5 A_{11}^q
     + \frac{1}{M} \varepsilon^{\mu\nu\rho\sigma} \gamma_{\mu} P_{\nu} k_{\rho} S_{\sigma}  A_{12}^q \,,
\end{align}

The amplitudes $A_i^q=A_i^q(P\cdot k,k^2)$ in (\ref{Eq:correlator-decompose}) are 
real functions of the Lorentz scalars $P\cdot k$ 
and $k^2$ \cite{Mulders:1995dh,Boer:1997nt}. 
The amplitudes $A_i^q$ are chiral-even for $i=2,\,3,\,6,\,7,\,8,\,12$ 
and chiral-odd for $i=1,\,4,\,5,\,9,\,10,\,11$. 
In QCD and in models with gauge field degrees of freedom,
in the definition of the quark correlator  (\ref{Eq:correlator-q})
Wilson lines must be included which run along a nearly light-like 
4-vector $n^\mu$ dictated by hard-momentum flow in the considered
process \cite{Collins-book}. The presence of the additional vector 
$n^\mu$ allows for 20 further Lorentz structures which are often denoted 
as $B_i$-amplitudes \cite{Kundu:2001pk,Goeke:2003az,Goeke:2005hb}.
The T-odd amplitudes $A_i^q$ for $i = 4,\,5,\,12$ vanish in quark models 
as do the pertinent T-odd TMDs because their modeling requires explicit 
gauge field degrees of freedom \cite{Pobylitsa:2002fr}. The T-odd
amplitudes are included in (\ref{Eq:correlator-decompose}) merely for 
completeness. In this work, we will focus on T-even TMDs,
see Appendix~\ref{App-A} for the explicit expressions.

\subsection{Quark correlator in the CPM}
\label{Sec-3:ampl-q}

In the CPM, one can explore the equation of motion for the quark fields 
$(i\slashed{\partial}-m_q)\,\Psi^q(z)=0$ in order to derive the following 
results for the amplitudes \cite{DAlesio:2009cps,Aslan:2022wqc} 
\ba
        A_1^q = \frac{m_q}{M}\,A_3^q, &&
        A_2^q = 0, \quad
        A_4^q = 0, \quad
        A_5^q = 0, \quad
        A_6^q =  \frac{m_q}{M}\,A_{10}^q, \nonumber\\
        A_7^q = -\frac{m_q}{M}\,A_{11}^q, &&
        A_9^q = 0, \quad 
        A_{10}^q = \frac{(P\cdot k)}{M^2}\,A_{11}^q-\frac{m_q}{M}\,A_8^q \,,\quad
        A_{12}^q = 0. 
        \label{Eq:amplitudes-CPM}
\ea
The T-odd amplitudes $A_4^q$, $A_5^q$, $A_{12}^q$ 
vanish in the CPM which is a general quark model
prediction due to the absence of gauge field 
degrees of freedom \cite{Pobylitsa:2002fr}. 
Interestingly, also the T-even amplitudes $A_2^q$ and 
$A_9^q$ vanish which is a specific feature of the CPM, 
and is in general not the case in other quark models. 
The amplitudes $A_1^q$, $A_6^q$, $A_7^q$ are proportional
to current quark masses and hence negligibly small for 
the light quark flavors. 
At this stage the relations (\ref{Eq:amplitudes-CPM})
imply that in the CPM 3 independent amplitudes exist 
which can be chosen to be the unpolarized amplitude
$A_3^q$, the chiral-even polarized  amplitude $A_8^q$, 
and the chiral-odd polarized amplitude $A_{11}^q$.

As shown in \cite{Aslan:2022wqc} one has two choices 
to treat quark polarization effects.
First, one can work with a CPM with 3 independent 
amplitudes \cite{DAlesio:2009cps} which corresponds to 
quarks in a mixed-spin state (as long as the inequality
$|A_8^q| < |A_{11}^q|$ is valid) \cite{Aslan:2022wqc}.
Alternatively, one can put the quarks in a pure-spin 
state in which case $|A_8^q| = |A_{11}^q|$ with the
physical-sign solution corresponding to $A_8^q=-A_{11}^q$ 
determined from a comparison to other models and lattice
QCD studies~\cite{Aslan:2022wqc}. This corresponds to a CPM
with 2 independent amplitudes which can be chosen to be 
the unpolarized amplitude $A_3^q$ and the polarized amplitude
$A_8^q$. The two choices can be summarized as follows
\ba
    \mbox{mixed-spin state CPM} & \Leftrightarrow &
    \mbox{3 linearly independent amplitudes:} 
    \quad A_3^q, \quad A_8^q, \quad A_{11}^q , \phantom{\frac11}
    \nonumber\\
    \mbox{pure-spin state CPM} & \Leftrightarrow &
    \mbox{2 linearly independent amplitudes:}  
    \quad A_3^q, \quad A_8^q = - A_{11}^q . 
    \label{Eq:CPM-mixed-pure}
\ea

\subsection{TMDs in the mixed-spin state version of the CPM}

The starting point for our investigation is the
CPM with quarks in the mixed-spin state. In this
section, we quote the results for T-even TMDs starting with the model expressions for unpolarized TMDs  
(we define $k^\pm=\frac{1}{\sqrt{2}}(k^0\pm k^3)$)
\bsub\label{Eq:rel-unp-q}
\ba
    f_1^q(x,k_T) &=& 
        2P^+\int dk^- \biggl[xA_3^q(P\cdot k,k^2)\biggr]_{k^+=xP^+},
        \label{Eq:f1-q-final}\\
    f^{\perp q}(x,k_T) &=& 
        2P^+\int dk^- \biggl[A_3^q(P\cdot k,k^2)\biggr]_{k^+=xP^+},\\
    e^q(x,k_T) &=& 
        2P^+\int dk^- \biggl[\frac{m_q}{M}\,
        A_3^q(P\cdot k,k^2)\biggr]_{k^+=xP^+}.
\ea\esub
The expressions for  chiral-even polarized TMDs in the 
mixed-spin state parton model are given by 
\bsub\label{Eq:mixed-spin-TMDq-all}
\ba
    g_1^q(x,k_T) &=& 2P^+\int dk^-\Biggl[
    \frac{x^2M^2 -x \,P\cdot k+m_q^2}{M^2} \,A_8^q(P\cdot k,k^2)
    -\frac{m_q}{M}\,x\,
    A_{11}^q(P\cdot k,k^2)\Biggr]_{k^+=xP^+}, 
    \label{Eq:mixed-spin-TMDq-g1}\\
    g_{1T}^{\perp q}(x,k_T) &=& 2P^+\int dk^- 
    \Biggl[xA_8^q(P\cdot k,k^2)-\dfrac{m_q}{M}
    A_{11}^q(P\cdot k,k^2)\Biggr]_{k^+=xP^+},
    \label{Eq:mixed-spin-TMDq-g1Tperp}\\
    g_T^q(x,k_T) &=& 2P^+\int dk^- \Biggl[
    \dfrac{\vec{k}_T^2+2m_q^2}{2M^2}\,A_8^q(P\cdot k,k^2)
    -\dfrac{m_q}{M}\;\dfrac{P\cdot k}{M^2}\,A_{11}^q(P\cdot k,k^2)
    \Biggr]_{k^+=xP^+},\label{Eq:mixed-spin-TMDq-gT}\\
    g_L^{\perp q}(x,k_T) &=& 2P^+\int dk^- \Biggl[
    \frac{x\,M^2-P\cdot k}{M^2}\,A_8^q(P\cdot k,k^2)\Biggr]_{k^+=xP^+},
    \label{Eq:mixed-spin-TMDq-gLperp}\\
    g_T^{\perp q}(x,k_T) &=& 2P^+\int dk^- \Biggl[A_8^q(P\cdot k,k^2)\Biggr]_{k^+=xP^+}.
    \label{Eq:mixed-spin-TMDq-gTperp}
\ea\esub
Finally, the model expressions for chiral-odd polarized 
TMDs are given by 
\bsub\begin{eqnarray}
    h_1^q(x,k_T) &=& 2P^+\int dk^- \Biggl[
    \dfrac{\vec{k}_T^2-2\,x\,P\cdot k}{2M^2}\,A_{11}^q(P\cdot k,k^2)
    +x\,\dfrac{m_q}{M}\,A_8^q(P\cdot k,k^2) \Biggr]_{k^+=xP^+},
    \label{Eq:mixed-spin-TMDq-h1}\\
    h_{1L}^{\perp q}(x,k_T) &=& 2P^+\int dk^- \Biggl[
    xA_{11}^q(P\cdot k,k^2) -\dfrac{m_q}{M}A_8^q(P\cdot k,k^2)\Biggr]_{k^+=xP^+},
    \label{Eq:mixed-spin-TMDq-h1Lperp}\\
    h_{1T}^{\perp q}(x,k_T) &=& 2P^+\int dk^- \Biggl[
    A_{11}^q(P\cdot k,k^2)\Biggr]_{k^+=xP^+},
    \label{Eq:mixed-spin-TMDq-h1Tperp}\\
    h_L^q(x,k_T) &=& 2P^+\int dk^- \Biggl[
    \frac{x^2M^2-2\,x\,P\cdot k}{M^2}\,A_{11}^q(P\cdot k,k^2)
    +\dfrac{m_q}{M}\;\dfrac{P\cdot k}{M^2}\,A_8^q(P\cdot k,k^2)
    \Biggr]_{k^+=xP^+},
    \label{Eq:mixed-spin-TMDq-hL}\\
    h_T^q(x,k_T) &=& 2P^+\int dk^- \Biggl[
    \frac{x\,M^2-P\cdot k}{M^2}\,A_{11}^q(P\cdot k,k^2)\Biggr]_{k^+=xP^+},
    \label{Eq:mixed-spin-TMDq-hT}\\
    h_T^{\perp q}(x,k_T) &=& 
    2P^+\int dk^- \Biggl[-\dfrac{P\cdot k}{M^2} A_{11}^q(P\cdot k,k^2)+\dfrac{m_q}{M}A_8^q(P\cdot k,k^2)\Biggr]_{k^+=xP^+}
    \label{Eq:mixed-spin-TMDq-hTperp}.
\end{eqnarray}\esub
For massless quarks in the mixed-spin state version 
of the CPM, the chiral-even (chiral-odd) polarized TMDs
are given entirely in terms of the chiral-even
(chiral-odd) amplitude $A_8^q$ ($A_{11}^q$). 

\subsection{Onshellness and a useful identity}

In the CPM the quarks are on-shell, i.e.\ the amplitudes 
$A_i^q(P\cdot k,k^2)$ are actually functions of the type
\be\label{Eq:mass-shell}
    A_i^q(P\cdot k,k^2) = F^q_i(P\cdot k)\,\delta(k^2-m_q^2)\,.
\ee
The explicit expressions for the functions 
$F_i^q(P\cdot k)$ can be found in
Refs.~\cite{Bastami:2020rxn,DAlesio:2009cps,Aslan:2022wqc}
and will not be needed in this work. However, we will need an
identity among the kinematic variables which holds under
the $k^-$ integration, and can be derived as follows.
Obviously, due to (\ref{Eq:mass-shell}) we have 
\ba
    0 
    =   \int dk^- \biggl(k^2-m_q^2\biggr) 
        A_i^q(P\cdot k,k^2) \Biggr|_{k^+=xP^+}
    =   \int dk^- \biggl(2k^+k^- - \vec{k}_T^2-m_q^2\biggr) 
        A_i^q(P\cdot k,k^2) \Biggr|_{k^+=xP^+}\nonumber\,.
\ea
Next, we notice that $2k^+k^-=2xP^+k^-=2xP\cdot k-2xP^-k^+
=2xP\cdot k-2x^2P^+P^-=2xP\cdot k-x^2M^2$ holds
under the integral where $k^+=xP^+$.
Inserting this in the above intermediate step,  dividing 
by $2x$ and rearranging, we obtain
\ba\label{Eq:useful-identity}
    \int dk^- \biggl(P\cdot k\biggr) 
        A_i^q(P\cdot k,k^2) \Biggr|_{k^+=xP^+}
    =   \int dk^- \biggl(
        \frac{x^2M^2+\vec{k}_T^2+m_q^2}{2x}\biggr) 
        A_i^q(P\cdot k,k^2) \Biggr|_{k^+=xP^+}\,.
\ea
Thus we see that under the $k^-$ integral due to the mass-shell condition
implicit in the amplitudes, cf.\ Eq.~(\ref{Eq:mass-shell}), we can 
replace the variable $P\cdot k$ by an expression determined in terms 
of $x$, $k_T$ and the nucleon and quark masses. This identity will
be helpful in the following.

\section{Consequences of imposing QMR\lowercase{s} in mixed-spin state CPM}
\label{Sec-4:imposing-QMRs}

Before we investigate the QMRs in the CPM, it 
is instructive to discuss first the example of the
qLIR in Eq.~(\ref{Eq:qLIR-example}). Here and in the following
it is convenient to reformulate the relations such that all TMDs 
appear on one side of the equation. Inserting the model expressions
(\ref{Eq:mixed-spin-TMDq-h1Lperp}, \ref{Eq:mixed-spin-TMDq-hT}, \ref{Eq:mixed-spin-TMDq-hTperp}) 
for $h_T^q$, $h_T^{\perp q}$, $h_{1L}^{\perp q}$
we obtain
\begin{align}
    h_T^{ q}(x,k_T)
    &
    -h_T^{\perp  q}(x,k_T)-
    h_{1L}^{\perp  q}(x,k_T)  
    =
    2P^+\int dk^- \Biggl[ 
    \frac{x\,M^2-P\cdot k}{M^2}\,A_{11}^q(P\cdot k,k^2)\nonumber\\
    &
    -\Bigl(-\dfrac{P\cdot k}{M^2} A_{11}^q(P\cdot k,k^2)  
    +\dfrac{m_q}{M}A_8^q(P\cdot k,k^2)\Bigr)
    -
    \Bigl(xA_{11}^q(P\cdot k,k^2) -\dfrac{m_q}{M}A_8^q(P\cdot k,k^2) \Bigr)
    \Biggr]_{k^+=xP^+} 
    = 0. \nonumber
\end{align}
We see that the qLIR (\ref{Eq:qLIR-example}) is 
valid for any $A_8^q(P\cdot k,k^2)$ and
$A_{11}^q(P\cdot k,k^2)$. This was to be expected.
The qLIRs only require the absence of gauge field
degrees of freedom, and thus must be valid in every 
quark model respecting Lorentz invariance. The
investigation of this and other qLIRs is a useful 
cross check for the theoretical consistency of a
model, but does not yield new insights. 
In this respect, the QMRs are more insightful 
as we shall discuss next.

\subsection{QMR between gear-worm functions, Eq.~(\ref{Eq:qm-rel-1})}

The TMDs $g_{1T}^{\perp  q}(x,k_T)$ and
$h_{1L}^{\perp  q}(x,k_T)$ 
are sometimes called gear-worm functions. 
In the spectator model study of 
Ref.~\cite{Jakob:1997wg} the QMR (\ref{Eq:qm-rel-1})
between these TMDs was derived which was later
confirmed in several other quark models.
Inserting the CPM expressions 
(\ref{Eq:mixed-spin-TMDq-g1Tperp}) for 
$g_{1T}^{\perp  q}(x,k_T)$ and
(\ref{Eq:mixed-spin-TMDq-h1Lperp}) for 
$h_{1L}^{\perp  q}(x,k_T)$,
the relation (\ref{Eq:qm-rel-1}) 
can be expressed as
\be
       g_{1T}^{\perp  q}(x,k_T) + h_{1L}^{\perp  q}(x,k_T)
       = 
       2P^+\int dk^- \Bigl(x-\dfrac{m_q}{M}\Bigr)
       \Biggl[A_8^q(P\cdot k,k^2)+A_{11}^q(P\cdot k,k^2)\Biggr]_{k^+=xP^+}.
\ee
Clearly, in the mixed-spin state version of the CPM, 
where the amplitudes $A_8^q$ and $A_{11}^q$ are
unrelated, the relation (\ref{Eq:qm-rel-1}) is 
not valid. 
If we would like the CPM to comply with this QMR,
then this is possible if and only if we impose 
the condition $A_8^q = - A_{11}^q$ 
which corresponds to the pure-spin state version 
of the CPM, cf.\ Eq.~(\ref{Eq:CPM-mixed-pure}).

\subsection{\boldmath QMR between $g_{T}^{\perp  q}$ 
and $h_{1T}^{\perp  q}$, Eq.~(\ref{Eq:qm-rel-2})}

The QMR (\ref{Eq:qm-rel-2}) connecting the TMDs
$g_{T}^{\perp  q}$ and $h_{1T}^{\perp  q}$ was,
to the best of our knowledge, first discussed in
Ref.~\cite{Avakian:2009jt}. Inserting CPM expressions
(\ref{Eq:mixed-spin-TMDq-gTperp}) and
(\ref{Eq:mixed-spin-TMDq-h1Tperp}) for 
$g_{T}^{\perp  q}(x,k_T)$ and 
$h_{1T}^{\perp  q} (x,k_T)$ respectively 
into Eq.~(\ref{Eq:qm-rel-2}) yields
\ba
    g_{T}^{\perp  q}(x,k_T) + h_{1T}^{\perp  q} (x,k_T) &=& 2P^+\int dk^- \Biggl[A_8^q(P\cdot k,k^2) + A_{11} ^q(P\cdot k,k^2)\Biggr]_{k^+=x P^+} \,.
\ea
Again we see that if the amplitudes $A_8^q$ 
and $A_{11}^q$ are unrelated, then the QMR 
(\ref{Eq:qm-rel-2}) is not valid which is the
case in the mixed-spin state version of the 
model. For the CPM to comply with this QMR,
it is necessary to introduce the condition
$A_8^q = - A_{11}^q$ which brings us to the 
pure-spin state version of the CPM, cf.\
Eq.~(\ref{Eq:CPM-mixed-pure}).

\subsection{\boldmath QMR between $g_L^{\perp q}$ 
and $h_T^q$, Eq.~(\ref{Eq:qm-rel-3})}

The QMR (\ref{Eq:qm-rel-3}) connecting the
twist-3 TMDs $g_{L}^{\perp q}$ and $h_{T}^q$ 
was derived for the first time in
Ref.~\cite{Jakob:1997wg} and later confirmed in
other models. 
Inserting respectively the model expressions 
(\ref{Eq:mixed-spin-TMDq-gLperp}) and 
(\ref{Eq:mixed-spin-TMDq-hTperp}) for the TMDs
$g_{L}^{\perp  q}(x,k_T)$ and $h_{T}^{  q}(x,k_T)$
into Eq.~(\ref{Eq:qm-rel-3}) leads immediately to
\ba
g_{L}^{\perp  q}(x,k_T)+h_{T}^{  q}(x,k_T)=2P^+\int dk^- 
    \left( \frac{x\,M^2-P\cdot k}{M^2}\right)\,\Biggl[A_8^q(P\cdot k,k^2)+A_{11}^q(P\cdot k,k^2)\Biggr]_{k^+=xP^+}\,.
\ea
Also in this case we see that in the mixed-spin 
state version of the CPM the relation
(\ref{Eq:qm-rel-3}) is not valid, unless we 
demand that $A_8^q = - A_{11}^q$ which is 
equivalent to introducing the pure-spin state 
version of the CPM,  cf.\ Eq.~(\ref{Eq:CPM-mixed-pure}).

\subsection{\boldmath QMR of helicity, transversity 
and pretzelosity, Eq.~(\ref{Eq:qm-rel-4})}

This QMR was, to the best of our 
knowledge, first discussed in \cite{Avakian:2008dz}.
The difference of $g_1^q$ and $h_1^q$ was 
known to be related in models to quark orbital
angular momentum \cite{Ma:1997gy,Ma:1998ar}
implying that pretzelosity is related
to quark orbital angular momentum \cite{She:2009jq}.
Although only a model relation, this is the only
connection of quark orbital angular momentum to 
TMDs known so far, and attracted a lot of interest.
The QMR (\ref{Eq:qm-rel-4}) and its connection 
to quark orbital angular momentum have been 
confirmed in several other model studies.
Inserting the model expressions
(\ref{Eq:mixed-spin-TMDq-g1},~\ref{Eq:mixed-spin-TMDq-h1},~\ref{Eq:mixed-spin-TMDq-h1Tperp})
into Eq.~(\ref{Eq:qm-rel-4}) we obtain the 
lengthy expression
\begin{align}
    g_{1}^{ q}(x,k_T) - h_{1}^{  q}(x,k_T) - h_{1T}^{\perp (1)q}(x,k_T)
    =  
    2P^+\int dk^- \Bigg[ 
    \Big( \dfrac{x^2M^2-x P\cdot k+m_q^2-xm_qM}{M^2}\Big)A_8^q(P\cdot k,k^2)
    &\nonumber\\
    + 
    \Big(\dfrac{xP\cdot k-xm_qM-k_T^2}{M^2}\Big)A_{11}^q(P\cdot k,k^2)
    &\Bigg]_{k^+=xP^+} \, . \nonumber
\end{align}
In order to proceed, we eliminate $P\cdot k$
under the integral by means of the identity
(\ref{Eq:useful-identity}). After rearranging,
the result can be expressed as
\ba
    g_{1}^{ q}(x,k_T) - h_{1}^{  q}(x,k_T) - h_{1T}^{\perp (1)q}(x,k_T)
    &=& 
    2P^+\int dk^- \dfrac{(xM-m_q)^2-k_T^2}{2M^2}\Bigg[A_8^q(P\cdot k,k^2)+ A_{11}^q(P\cdot k,k^2)\Bigg]_{k^+=xP^+}. 
\ea
As in the previous cases, we see that in the
mixed-spin state version of the CPM the relation
(\ref{Eq:qm-rel-4}) is not supported. For this
QMR to be valid in the CPM we must introduce the
condition $A_8^q = - A_{11}^q$ which is equivalent
to introducing the pure-spin state version of the
CPM, cf.\ Eq.~(\ref{Eq:CPM-mixed-pure}).

\subsection{\boldmath QMR of twist-3 TMDs $g_T^q$ 
and $h_L^q$ to pretzelosity, Eq.~(\ref{Eq:qm-rel-5})}

We now turn our attention to the last linear QMR 
which connects $g_T^{ q}(x,k_T)$, $h_L^{ q}(x,k_T)$ 
and the transverse moment of pretzelosity. 
Inserting in (\ref{Eq:qm-rel-5})
the model expressions 
(\ref{Eq:mixed-spin-TMDq-gT},
\ref{Eq:mixed-spin-TMDq-h1Tperp},
\ref{Eq:mixed-spin-TMDq-hL}) for
$g_T^q(x,k_T)$, $h_L^q(x,k_T)$ and 
$h_{1T}^{\perp q}(x,k_T)$ gives
\be
    g_{T}^{ q}(x,k_T) - h_{L}^{  q}(x,k_T) - h_{1T}^{\perp (1)q}(x,k_T)
    =
    2P^+\int dk^- \left(\dfrac{2 m_q^2 +k_T^2}{2M^2}-\dfrac{ m_q }{M}\dfrac{P \cdot k}{M^2}\right)\Bigg[A_8^q(P\cdot k,k^2)+ A_{11}^q(P\cdot k,k^2)\Bigg]_{k^+=xP^+}. 
\ee
As in the previous cases, we observe that the 
QMR (\ref{Eq:qm-rel-5}) is not valid in the 
mixed-spin state version of the CPM, and can
be satisfied only when one introduces the 
condition $A_8^q = -A_{11}^q$, i.e.\
the pure-spin state version of the model.

\subsection{\boldmath Nonlinear QMR between 
$h_1^q$, $h_{1L}^{\perp q}$, $h_{1T}^{\perp q}$
in Eq.~(\ref{Eq:qm-rel-7})}
\label{subsec-QMR-3a}

The nonlinear QMR (\ref{Eq:qm-rel-7}) was derived in
Ref.~\cite{Avakian:2010br}. Inserting the model expressions
(\ref{Eq:mixed-spin-TMDq-h1},~\ref{Eq:mixed-spin-TMDq-h1Lperp},~\ref{Eq:mixed-spin-TMDq-h1Tperp}) 
into the nonlinear relation (\ref{Eq:qm-rel-7}) we obtain
\begin{align}
    2\,h_1^q(x,k_T)\,h_{1T}^{\perp q}(x,k_T) +
    h_{1L}^{\perp q}&(x,k_T)^2 
     \nonumber\\
    =
    (2P^+)^2\iint dk^-d{k'}^- 
    &\Biggl[
    x\,\frac{m_q}{M}\Biggl(
         A_8^q(P\cdot k,k^2)A_{11}^q(P\cdot k',k'{}^2)
        -A_8^q(P\cdot k',k'{}^2)A_{11}^q(P\cdot k,k^2)\Biggr) \nonumber\\
    &+\frac{m_q^2}{M^2}\Biggl(
         A_8^q(P\cdot k,k^2)A_8^q(P\cdot k',k'{}^2)-
         A_{11}^q(P\cdot k,k^2)A_{11}^q(P\cdot k',k'{}^2)\Biggr)
    \Biggr]_{k^+=xP^+}, \label{Eq:der-non-lin-1a}
    \end{align}
where $k=(k^+,k^-,\vec{k}_T)$ 
and $k'=(k^+,k'{}^-,\vec{k}_T)$ and we used the identity
(\ref{Eq:useful-identity}) to eliminate the variable 
$P\cdot k$ under the integral. 
In order to proceed, we repeat the calculation leading to
(\ref{Eq:der-non-lin-1a}) with the dummy integration
variables $k^-$ and ${k'}^-$ interchanged, and take 
the average of the two results. 
In this way, the "mixed terms" in the first term on 
the right-hand side of (\ref{Eq:der-non-lin-1a}) 
with $A_8^qA_{11}^q$ cancel out, and we obtain
\begin{align}
    2\,h_1^q(x,k_T)\,h_{1T}^{\perp q}(x,k_T) +
    h_{1L}^{\perp q}&(x,k_T)^2 
    \nonumber\\
    =
    (2P^+)^2\iint dk^-d{k'}^-
    &    \frac{m_q^2}{M^2}\Biggl[
         A_8^q(P\cdot k,k^2)A_8^q(P\cdot k',k'{}^2)-
         A_{11}^q(P\cdot k,k^2)A_{11}^q(P\cdot k',k'{}^2)
    \Biggr]_{k^+=xP^+}, \label{Eq:der-non-lin-1b}
    \end{align}
It is convenient to rewrite this result in the following
equivalent way
\begin{align}
    2\,h_1^q(x,k_T)\,h_{1T}^{\perp q}(x,k_T) +
    h_{1L}^{\perp q}&(x,k_T)^2 
    \nonumber\\
    =
    (2P^+)^2\iint dk^-d{k'}^- 
    &\frac{m_q^2}{M^2}\Biggl[
         \Biggl(A_8^q(P\cdot k,k^2)+A_{11}^q(P\cdot k,k^2)\Biggr)
         \Biggl(A_8^q(P\cdot k',k'{}^2)-A_{11}^q(P\cdot k',k'{}^2)\Biggr)
         \Biggr]_{k^+=xP^+}. \label{Eq:der-non-lin-1c}
    \end{align}
In order to show that (\ref{Eq:der-non-lin-1c})
is equivalent to (\ref{Eq:der-non-lin-1b}) one can
apply the trick with repeating the calculation with 
the dummy integration variables $k^-$ and $k'{}^-$
interchanged, and taking the average.

As in the case of linear QMRs, the nonlinear relation
(\ref{Eq:qm-rel-7}) is in general not valid in the CPM
version with quarks in a mixed-spin state. Interestingly
and in contrast to the linear case, the violation of the
nonlinear QMR (\ref{Eq:qm-rel-7}) is, however, a small
effect proportional to $m_q^2/M^2$ which is 
numerically of the order ${\cal O}(10^{-6})$
for the light $u$- and $d$-flavors. This observation
may have interesting consequences on which we shall
comment in Sec.~\ref{Sec-5:discussion}.

If we insist
on the nonlinear QMR (\ref{Eq:qm-rel-7}) to be 
exactly valid for $m_q\neq0$, then we see from the final expression
(\ref{Eq:der-non-lin-1c}) that there are two solutions:
$A_8^q=\pm A_{11}^q$. It is not surprising to find two
solutions, as we deal with a quadratic equation. 
Both solutions were encountered in \cite{Aslan:2022wqc},
and $A_8^q= + A_{11}^q$ was recognized to be an 
unphysical solution as it would imply opposite signs for
quark helicity and transversity TMDs in contradiction 
to results from other models and lattice QCD.
The solution $A_8^q=- A_{11}^q$ leads to like signs 
for quark helicity and transversity TMDs in agreement 
with other models and lattice QCD and constitutes 
therefore the physical solution \cite{Aslan:2022wqc}.
Thus, the CPM with massive quarks complies exactly
with the nonlinear QMR (\ref{Eq:qm-rel-7}) if and 
only if we use the pure-spin version of the model.

\subsection{\boldmath Nonlinear QMR between $g_{1T}^{\perp q}$, $g_L^{\perp q}$, $g_T^q$, $g_{T}^{\perp q}$ in Eq.~(\ref{Eq:qm-rel-8})}

The nonlinear QMR (\ref{Eq:qm-rel-8}) was also derived in
Ref.~\cite{Avakian:2010br}. Inserting the model expressions
(\ref{Eq:mixed-spin-TMDq-g1Tperp},~\ref{Eq:mixed-spin-TMDq-gT},~\ref{Eq:mixed-spin-TMDq-gLperp},~\ref{Eq:mixed-spin-TMDq-gTperp}) 
into the nonlinear relation (\ref{Eq:qm-rel-8}) we obtain
\begin{align}
      2\,g_{1T}^{\perp q}(x,k_T)\,g_L^{\perp q}(x,k_T) +
    & 2\,g_T^q(x,k_T)\,g_T^{\perp q}(x,k_T)
      - g_{1T}^{\perp q}(x,k_T)^2 
    \nonumber\\
    =
    (2P^+)^2\iint dk^-d{k'}^- 
    &\Biggl[
    x\,\frac{m_q}{M}\Biggl(
         A_8^q(P\cdot k,k^2)A_{11}^q(P\cdot k',k'{}^2)
        -A_8^q(P\cdot k',k'{}^2)A_{11}^q(P\cdot k,k^2)\Biggr) \nonumber\\
    &+\frac{m_q^2}{M^2}\Biggl(
         A_8^q(P\cdot k,k^2)A_8^q(P\cdot k',k'{}^2)-
         A_{11}^q(P\cdot k,k^2)A_{11}^q(P\cdot k',k'{}^2)\Biggr)
    \Biggr]_{k^+=xP^+}, \label{Eq:der-non-lin-2a}
    \end{align}
where we used the identity (\ref{Eq:useful-identity})
to eliminate the variable $P\cdot k$ in the coefficient
of the $A_8^q(k\cdot P,k^2)A_8^q(P\cdot k',k'{}^2)$-term (in other cases $P\cdot k$ 
cancels out). The expression under the 
integral of (\ref{Eq:der-non-lin-2a}) coincides
with the expression in (\ref{Eq:der-non-lin-1a})
and the further steps  continue from here in the 
same way as in Sec.~\ref{subsec-QMR-3a} 
including all considerations and conclusions.

\section{Discussion of the results}
\label{Sec-5:discussion}

In order to better understand the physical 
implications of our results it is instructive to 
briefly review the relation between the two 
versions of the CPM  \cite{Aslan:2022wqc}.
In Ref.~\cite{Aslan:2022wqc} it was recognized
that for massive quarks, $m_q\neq0$, the quark
correlator can be expressed compactly be introducing
an axial vector $w_q^\mu$ which has the properties of 
a quark polarization vector and satisfies 
$k\cdot w_q=0$. One then 
has\footnote{\label{foot-1} The
    polarization of massless quarks cannot be 
    described in terms of a polarization vector. 
    But ultimately in the massless case, one has the
    same choice of putting a quark in a pure-spin vs
    mixed-spin state. In this work, it is more
    insightful to work with the more general case
    $m_q\neq 0$. But if desired, the current quark
    mass effects can be neglected at any stage, 
    cf.\ \cite{Aslan:2022wqc}.}
a choice:
a quark can be in a pure-spin state 
with $w_q^2=-1$, or mixed-spin state 
with $-1 < w_q^2 < 0$. These two choices
lead to the two versions of the model, 
cf.\ Eq.~(\ref{Eq:CPM-mixed-pure}).

It is an interesting question which of the 
two CPM versions might be more realistic
from phenomenological point of view.
At first glance, one could suspect the
mixed-spin state version of the CPM to be
phenomenologically more realistic due to 
a larger flexibility with three independent
covariant functions which can be uniquely 
determined from parametrizations 
of unpolarized, helicity, and 
transversity parton distribution functions \cite{Harland-Lang:2014zoa,deFlorian:2014yva,Kang:2015msa,Sato:2016tuz,Radici:2018iag,Bailey:2020ooq,Cammarota:2020qcw}.
This question can be answered by future
studies, when more of the TMDs become
better known and constrained by data.

Meanwhile, one could also try to address
this question based on what is known about TMDs from 
other quark models. A striking observation is that
a large class of quark models supports the QMRs.
Thus, one could wonder whether, based on a comparison 
to other models, for instance the linear QMRs 
(\ref{Eq:quark-model-relations-linear})
should also hold in the CPM. 
If one would like the CPM to comply with the QMRs, 
then one must introduce a condition between the 
polarized amplitudes, namely $A_8^q = -A_{11}^q$
as shown in Sec.~\ref{Sec-4:imposing-QMRs},
which leads at once to the pure-spin state 
version of the CPM. 

To be more precise, when one approaches the issue from
the point of view of a quark polarization vector
$w_q^\mu$, the pure-spin condition $w_q^2=-1$
only tells us that $|A_8^q| = |A_{11}^q|$, and the 
CPM {\it per se} is not able to predict the sign of 
the chiral-odd TMDs. It is necessary to resort to 
results from other models and lattice QCD to 
determine the physical solution \cite{Aslan:2022wqc}. 
Here the situation is different. The linear QMRs
already ``encode'' the information from other models
about the relative signs of the polarized chiral-even
and chiral-odd TMDs. By imposing the linear QMRs in
the mixed-spin version of the CPM, one is
unambiguously lead to the condition 
$A_8^q = -A_{11}^q$ without encountering any
spurious unphysical solution.
 
Thus, there are two ways to introduce the CPM 
with quarks in a pure-spin state: (a) by 
demanding that $w_q^2=-1$ and determining the
physical solution, or (b) by demanding that
the model be compliant with the QMRs observed 
in other quark models. The two procedures are
conceptually quite different, but 
nevertheless equivalent.
This is an interesting observation in itself, 
and gives new insights on the CPM. Notice that 
this observation is independent of whether one
considers massive quarks or neglects quark mass 
effects, cf.\ footnote~\ref{foot-1}.

These considerations are of interest beyond
the CPM and give rise to a question 
regarding the spin state of quarks in other 
models which, to the best of our knowledge, 
has not been addressed in literature. 
Considering that the TMDs in the CPM comply 
with QMRs if and only if the quarks are in
pure-spin state, one may wonder whether the
reverse is true: if a quark model supports 
the QMRs, are the quarks in this model 
necessarily in a pure-spin state? We do not 
know the answer. It will be interesting to 
address this question in other models.

The above remarks about the QMRs leading to
the condition $A_8^q = -A_{11}^q$ refer to the 
linear case (\ref{Eq:quark-model-relations-linear}).
For the non-linear QMRs 
(\ref{Eq:quark-model-relations-nonlinear}) 
the situation is different. These relations are
quadratic in TMDs, and hence it is not 
surprising to encounter two solutions 
$A_8^q = \pm A_{11}^q$ one of which is physical
and the other unphysical. As a consequence,
with non-linear QMRs
(\ref{Eq:quark-model-relations-nonlinear})
alone, we would need to use additional constraints 
to determine the physical and eliminate the 
unphysical solution ---
analogously as it was done with the two solutions of 
the condition $w_q^2=-1$ in \cite{Aslan:2022wqc}.

However, there is an interesting difference between 
the ways the CPM can comply with linear and 
non-linear QMRs which bears an unexpected observation.
In the more general mixed-spin state version
of the CPM, the violation of the non-linear QMRs
(\ref{Eq:quark-model-relations-nonlinear}) is
proportional to the square of
the current quark masses. In other words,
already in the mixed-spin version of the CPM
the non-linear QMRs
(\ref{Eq:quark-model-relations-nonlinear})
are supported modulo current quark mass effects
proportional to $m_q^2/M^2$ which is numerically 
an effect of order $10^{-6}$ for the up- and
down-flavors. 
 
This is an interesting observation for the following
reason. The description of TMDs in QCD
becomes equivalent to that in the parton model 
in the Wandzura-Wilczek-type (WW-type)
approximation \cite{Avakian:2007mv}.
This approximation consists in exploring the QCD 
equations of motion for twist-3 TMDs to relate them
to the better known twist-2 TMDs and the so-called 
tilde terms which are contributions due to
quark-gluon-quark matrix elements and current quark 
mass terms. Neglecting the tilde- and current quark
mass terms constitutes the WW-type approximation.
(The attribute ``type'' is added to distinguish 
the more complex TMD case from the original 
WW-approximation for the colinear function 
$g_T^q(x)$ \cite{Wandzura:1977qf}.)
 The exploration of the free equation of motion
 in the parton model generates exactly the same
 mass terms as in QCD but of course no tilde-terms.
 In this sense, the predictions of the parton model
 are equivalent to the description of TMDs in QCD
in the WW-type approximation.

The linear QMRs 
(\ref{Eq:quark-model-relations-linear})
hold in the CPM only if one introduces an
{\it additional} constraint which is equivalent 
to putting the quarks in a pure-spin state. It 
remains to be seen whether this leads to a realistic
modelling of the nucleon structure from 
phenomenological point of view. 
However, the non-linear QMRs 
(\ref{Eq:quark-model-relations-nonlinear})
do not require such an additional condition,
and are valid also for (light) quarks in a
mixed-spin state. 
This could imply that the non-linear QMRs 
(\ref{Eq:quark-model-relations-nonlinear})
are more likely to be supported in QCD
because no additional (pure-spin state)
condition is required for their validity.

The observation that the non-linear QMRs 
(\ref{Eq:quark-model-relations-nonlinear})
could be valid in the WW-type approximation
is interesting.
The WW-type approximation has been explored
for phenomenological applications for instance
in \cite{Bastami:2018xqd}. The 
quality of this approximation cannot be
determined a priori, and it needs to be 
investigated on a case by case basis 
because different quark-gluon-quark 
matrix elements are neglected in each case.
In some cases the WW-type approximation
was shown to work with a phenomenologically 
useful approximation 
\cite{Bastami:2018xqd,Bhattacharya:2021twu}
and in one case there is support from
lattice QCD \cite{Bhattacharya:2021moj}.
It will be very interesting to investigate
whether the non-linear QMRs 
(\ref{Eq:quark-model-relations-nonlinear})
could be valid in QCD with a similarly 
useful approximation

\section{Conclusions and outlook}
\label{Sec-6:conclusions}

In this work, we have investigated the 
quark-model relations (QMRs) in the mixed-spin
version of the covariant-parton model (CPM).
The equations of motion in the CPM imply
some conditions among the amplitudes in the
quark correlator, but leave the amplitudes
$A_8^q $ and $A_{11}^q$ unrelated. 
We have shown that the linear QMRs are not
valid, unless one introduces the condition 
$A_8^q = - A_{11}^q$. This condition is 
equivalent to putting the quarks in a pure-spin
state (more precisely: the  pure-spin state 
condition only implies $|A_8^q| = |A_{11}^q|$
and does not determine the relative sign).

Our results are of interest because they give 
insights on the CPM and raise interesting
questions about quark models and QMRs.
The observation that imposing linear QMRs
is equivalent to putting the quarks in a
pure-spin state is primarily an insight 
about the CPM but may be of interest also
beyond this model for the following reason. 
In the CPM the two statements, 
(i) quarks are in the pure-spin state and 
(ii) model complies with QMRs, are equivalent.
It will be interesting to investigate 
whether this is the case also in other models:
if a quark model obeys the QMRs, are then the
quarks in this model in a pure-spin state? 
This aspect deserves further investigations.

We also learn an interesting lesson about QMRs.
In QCD, each TMD is an independent function 
describing a different aspect of the nucleon
structure, and no relations among TMDs exist. 
In quark models, the situation can be simpler 
and relations among TMDs may exist. Such 
relations become particularly interesting if
they are supported by a wide class of different
quark models as is the case with the QMRs which
arise from a certain symmetry of the nucleon wave
function which is present in many 
(though not all) quark models. 

QMRs become even more interesting if they 
require only general model assumptions. In the CPM,
the linear QMRs require a strong model assumption,
namely the quarks must be in pure spin state. 
The situation is different for the nonlinear
QMRs. These relations become exact in the CPM
for quarks in a pure-spin state and/or for 
massless quarks.
However, even in the most general case in the CPM,
i.e.\ for massive quarks in a mixed-spin state,
the nonlinear QMRs are still valid to a very good 
approximation, namely up to negligibly
small quadratic quark mass effects 
$\propto m_q^2/M^2$.

Thus, the nonlinear QMRs are practically supported 
in the CPM independently of the quark spin state.
In other words, the nonlinear QMRs require no
strong model assumption
(like the pure-spin condition). 
From the point of view of the CPM, all that is 
required for the nonlinear QMRs is the absence
of interactions. From the point of view of QCD,
this in turn means that the non-linear QMRs 
could be valid in the WW-type approximation.

It is important to remark, that even if the QMRs
were valid at one scale, due to the
different evolution equations of the different
TMDs, they would not be valid at other scales. 
However, considering their crude nature, 
the "accuracy" of quark models can be 
expected to be around ${\cal O}$(30-40$\,\%$) 
\cite{Boffi:2009sh,Pasquini:2011tk}, the 
TMD evolution effects are not a dominant
uncertainty. It will be interesting to see 
whether phenomenological extractions or 
lattice QCD results will support, at some
scale, predictions from quark models like 
the CPM within such model accuracy.

The spin state of a quark in QCD is not easy to 
determine
\cite{Dalitz:1988aq,Efremov:1992pe,Collins:1992kk}.
The comparison of the CPM predictions to
phenomenological results for TMD extractions 
will constitute one way to infer to which extent
the quarks in QCD can be viewed as being in 
a pure- or mixed-spin state.  
It will be interesting to shed more light on 
the polarization state of the quarks in the 
nucleon based on dedicated phenomenological,
model and lattice QCD studies.

\ \\
\noindent
{\bf Acknowledgments.}
This work was partly supported by NSF 
under the Award No.~1812423 and  Award No.\ 2111490,
by the U.S.~Department of Energy, under the contract
no.~DE-AC05-06OR23177 under which Jefferson Science 
Associates, LLC operates Jefferson Lab, and within 
the framework of the TMD Collaboration.

\appendix

\section{Quark model expressions for T-even TMDs}
\label{App-A}

In this Appendix, to make this work self-contained,
we list the quark model expressions for T-even TMDs 
in terms of the amplitudes defined in
(\ref{Eq:correlator-decompose}). These expressions are 
valid in all models without gauge field degrees of freedom. 
In QCD, the TMDs depend on the renormalization scale $\mu^2$ 
and the scale $\zeta$ at which lightcone divergences are
regulated. 
In this work, we do not indicate the scales for brevity.
The determination of these scales in a model calculation is
is an important part of the modelling. In previous works in 
the CPM, the scales were assumed to be 
$\mu^2=\zeta \simeq = (\mbox{3-4})\,{\rm GeV}^2$.

In the twist-2 case, the expressions for TMDs read
\begin{eqnarray}
    \label{Eq:TMDq_f_1}
         f_1^q(x,k_T) \; &=& \; 
         2P^+ \int dk^- \biggl[A_2^q + x A_3^q\biggr]_{k^+=xP^+} \, , \\
    \label{Eq:TMDq_f^perp_1T}
         g_1^q(x,k_T) \; &=& \; 
         2P^+ \int dk^- 
        \biggl[ -A_6^q-\frac{P \cdot k-M^2x}{M^2}(A_7^q+xA_8^q)
        \biggr]_{k^+=xP^+}\, , \\
    \label{Eq:TMDq_g_1T}
         g_{1T}^{\perp q}(x,k_T) \; &=& \; 
         2P^+ \int dk^-\biggl[A_7^q+xA_8^q\biggr]_{k^+=xP^+} \, , \\
    \label{Eq:TMDq_h_1}
         h_1^q(x,k_T) \; &=& \; 
         2P^+ \int dk^- 
         \biggl[ -A_9^q-xA_{10}^q+\frac{\vec{k}_T^{\,2}}{2M^2} \; A_{11}^q 
         \biggr]_{k^+=xP^+} \, ,\\
    \label{Eq:TMDq_h^perp_1L}
         h^{\perp q}_{1L}(x,k_T) \; &=& \; 
         2P^+ \int dk^- 
         \biggl[ A_{10}^q - \frac{P \cdot k - M^2x}{M^2} \; A_{11}^q
         \biggr]_{k^+=xP^+} \, , \\
    \label{Eq:TMDq_h^perp_1T}
         h^{\perp q}_{1T}(x,k_T) \; &=& \; 
         2P^+ \int dk^- \biggl[ A_{11}^q \biggr]_{k^+=xP^+} \, .
\ea
In the twist-3 case, the expressions are given by
\ba
    \label{Eq:TMDq_e}
         e^q(x,k_T) \; &=& \; 
         2P^+ \int dk^- \biggl[ A_1^q \; \biggr]_{k^+=xP^+} \, , \\
    \label{Eq:TMDq_f^perp}
         f^{\perp q}(x,k_T) \; &=& \; 
         2P^+ \int dk^- \biggl[ A_3^q \; \biggr]_{k^+=xP^+} \, , \\
         \label{eq:appg_T}
         g_T^q(x,k_T) \; &=& \; 
         2P^+ \int dk^- 
         \biggl[ -A_6^q + \frac{\vec{k}_T^{\,2}}{2M^2} \; A_8^q
         \biggr]_{k^+=xP^+} \, , \\
    \label{Eq:TMDq_g^perp_L}
         g^{\perp q}_L(x,k_T) \; &=& \; 
         2P^+ \int dk^- 
         \biggl[ - \frac{P \cdot k-M^2x}{M^2}\; A_8^q
         \biggr]_{k^+=xP^+} \, , \\
    \label{Eq:TMDq_g^perp_T}
         g^{\perp q}_T(x,k_T) \; &=& \; 
         2P^+ \int dk^- \biggl[ A_8^q \; \biggr]_{k^+=xP^+}\, , \\
    \label{Eq:TMDq_h_L}
         h_L^q(x,k_T) \; &=& \; 
         2P^+ \int dk^- 
         \biggl[ -A_9^q-\frac{P \cdot k}{M^2} \; A_{10}^q 
         + \frac{(P \cdot k-M^2x)^2}{M^4} \, A_{11}^q 
         \biggr]_{k^+=xP^+} \, , \\
    \label{Eq:TMDq_h_T}
         h_T^q(x,k_T) \; &=& \; 
         2P^+ \int dk^- 
         \biggl[ - \frac{P \cdot k-M^2x}{M^2} \; A_{11}^q
         \biggr]_{k^+=xP^+} \, , \\
    \label{Eq:TMDq_h^perp_T}
         h^{\perp q}_T(x,k_T) \; &=& \; 
         2P^+ \int dk^- 
         \biggl[ -A_{10}^q \biggr]_{k^+=xP^+} \, . 
\end{eqnarray}
In QCD also $B_i^q$ amplitudes enter, see e.g.\
\cite{Metz:2008ib} for the full expressions. 
But in quark models the 14 T-even TMDs are expressed in terms of
9 T-even $A_i^q$ amplitudes. This implies 5 relations, namely the qLIRs mentioned in Sec.~\ref{Sec-2:QMRs}. 

We also remark that, in contrast to QCD, in the CPM 
no UV- or rapidity divergences appear. This allows 
one to relate TMDs and colinear parton
distribution functions simply as 
$f_1^q(x) = \int d^2k_Tf_1^q(x,k_T)$ with a finite
$k_T$-integration, which in QCD 
\cite{Collins-book} as well as in some models
\cite{Schweitzer:2012hh,Aslan:2021new2}
is spoiled by the appearance of divergences.

\end{document}